\documentclass[twocolumn,showpacs,preprintnumbers,amsmath,amssymb]{revtex4}

\usepackage{graphicx}% Include figure files
\usepackage{dcolumn}% Align table columns on decimal point
\usepackage{bm}% bold math
\usepackage{marginnote}
\usepackage{url}
\usepackage{color}

\usepackage{setspace}
\begin{document}

\preprint{xxxx}

\title{Can plasma physics establish a significant bound on long-range dark matter interactions?}

\author{K. Schoeffler$^{1,2}$}%
\email{Kevin.Schoeffler@rub.de}
\author{N. Shukla$^{1,3}$}
\email{n.shukla@cineca.it}

\author{L. O. Silva$^{1}$}%
\email{luis.silva@tecnico.ulisboa.pt}
\affiliation{$^{1}$GoLP/Instituto de Plasmas e Fus\~ao Nuclear, Instituto Superior T\'ecnico, Lisbon, Portugal}%
\affiliation{$^{2}$Institut f\"ur Theoretische Physik, Ruhr-Universit\"at Bochum, Bochum, Germany}%
\affiliation{$^{3}$CINECA High-Performance Computing Department, Casalecchio di Reno, Italy}%

\date{\today}% It is always \today, today,

%: Abstract
\begin{abstract}
	Dark matter has been theorized to be charged under its own ``dark electromagnetism" (dark-EM). Under this hypothesis, dark matter can behave like a cold collisionless plasma of self-interacting dark matter particles and exhibit plasma-like instabilities with observational consequences. Using the results published in~\cite{Shukla2022}, which studied the degree of slowdown between two interpenetrating $e^-\,e^+$ plasma clouds due to plasma instabilities, estimates of similar interactions for colliding ``dark plasmas" are explored. Comparison with astronomical observations reveals strong new constraints on dark-EM with the dark electromagnetic self-interaction $\alpha_{D} < 4 \times 10^{-25}$.

\end{abstract}

\pacs{xxxxxx}% PACS, the Physics and Astronomy
\maketitle

\section{Introduction}
%\emph{The mystery of dark matter (DM)\,:\textemdash} 
The nature of the dark matter, which comprises more than 80\% of the mass of the universe, remains a mystery \cite{DM1,DM2}. Having been observed only through its gravitational interactions, we know very little about the underlying particle physics of dark matter \cite{DM3,DM4,DM5,DM6,DM7}. A huge array of theoretical possibilities remains open, with a range of potential masses spanning 50 orders of magnitude from $\sim$$10^{-22}\,\mathrm{eV}$ to $\sim$$10^{19}\,\mathrm{GeV}$. Dark matter could have a wide variety of possible non-gravitational interactions both with itself and with other particles, although all we can currently say is that they must be weak enough to have evaded observation \cite{DM8,DM9}. Discovering such interactions would be a huge leap forward in understanding dark matter and is one of the biggest goals of modern particle physics.

%\emph{How do they interact?\,:\textemdash}
Interestingly, as we explore in this paper, the most minimal type of dark matter interaction can have rich astrophysical consequences over the entire allowed mass range. An unbroken U(1) gauge force, similar to electromagnetism but acting only between dark matter particles, would mediate long-range dark-matter self-interactions. We refer to such a force as {\it dark electromagnetism} (dark-EM) \cite{DM10,DM11}. This is a very natural possibility that arises in a wide range of underlying dark matter models \cite{DM6,DM12,DM22, DM23, DM24, Rennan}. The stability of dark matter constrains these models as the conservation of dark-EM charge could forbid dark matter decay. In general, constraints on this scenario or observable consequences are of broad interest and merit thorough study. Assuming the interactions are weak enough not to bind dark matter particles together, dark matter would then consist of a net-neutral plasma of dark-EM charged particles. In this paper, we investigate the effects this would have on halo dynamics.

%\emph{Dark-matter could interact due to the presence of dark-charge particles\,:\textemdash}
The dynamics of dark matter halos is the natural place to observe dark matter self-interactions. A recent observation of the ``Bullet Cluster'' (1E 0657-558), where a collision between a subcluster and the main cluster confirmed that the dark matter halo of the subcluster passes through the main cluster with no visible offset between the stars and the DM, thus indicating weak self-interactions. The seemingly clean passage of one halo through another in the ``Bullet Cluster'' led to well-known bounds being placed on $2\to2$ hard scattering (i.e., short-range interactions) \cite{DM25,DM26}. Dark-EM also gives rise to conceptually different effects, which result in far stronger bounds, due to the collective dynamics of the dark-matter plasma. The dark-EM interaction would act as a $N\to\,N$ rather than a $2\to2$ scattering process, where $N$ is extremely large. The bounds from $2\to2$ scattering require that a typical dark matter particle has never undergone a hard scattering \cite{DM12, DM13}. This indicates that the dark-matter plasma is in the ``collisionless'' regime \cite{DM14, DM15}. Under this hypothesis, we shall see whether the most minimal type of dark matter self-interaction via a new long-range force analogous to the electromagnetic interaction in the standard model could have an impact on DM self-interactions. This possibility raises the question of whether plasma-like collisionless instabilities may have a significant impact on galaxy and cluster dynamics. Resolving this will determine whether or not such an interaction is consistent with current observations, and the impact of plasma-like collisionless instabilities on galactic dynamics.

%\emph{Plasma instabilities\,:\textemdash}
%\emph{Pedagogical description of plasma instabilities known in literature\,:\textemdash}
%-----------------------------------Plasma instability-------------------------------------%
In this work, we consider the simple possibility of an electromagnetic self-interaction between two collisionless dark matter plasma slabs. The equations that govern DM self-interactions are identical to that of a collisionless $e^-, e^+$ plasma. Following the work done in reference~\cite{DM10}, we analyze simulations of interacting equal mass nonrelativistic plasmas from reference~\cite{Shukla2022} to draw our conclusions.
Collisionless plasma dynamics is both a well-studied field and an area of active research with rich dynamics that are still not fully understood. It is known that two counterpropagating plasmas are subjected to several microinstabilities that generate growth of electromagnetic fields, involving transverse and parallel modes. The full unstable wavenumber k spectrum has been intensively studied in the cold plasma limit \cite{DM16, DM17, DM18, DM19}. Three main dominating instabilities exist with different wave vectors with respect to the flow. First, the two-stream instability, which has a wave vector aligned with the flow, is driven by the two-peaked nature of the velocity distribution \cite{Bohm}. Second, anisotropy in the velocity spread in different directions (larger along the flow direction) excites the Weibel/current filamentation instability with a wave vector normal to the flow \cite{Weibel, Boella-2017, Muggli-2013, Sari-Nature-2015, shukla18}. Finally, a hybrid of these two modes, with a wave vector with an angle oblique to the flow, is known as the oblique instability \cite{DM16,DM18}.

Plasma instabilities are well understood analytically in the linearized regime with small perturbations to an infinite homogeneous plasma. Furthermore, theoretical estimates of the linear growth of electromagnetic field generated via plasma instabilities, complemented by numerical simulations, have been well studied during the interaction of two plasma slabs \cite{Luis-2002,Shukla-JPP-2012}.  The electromagnetic fields driven by these plasma instabilities will lead to bulk slowdown of the counterpropagating plasma slabs as long as there is sufficient time for the instabilities to grow. However, the exact time required before a significant slowdown occurs depends on nonlinear effects not included in such an analysis and can only be captured via numerical simulations, as performed in this paper. Our studies, therefore, can place a limit on the interaction strength by considering the full nonlinear dynamics associated with these instabilities.

%The dimensionless quantities, $\alpha_B = U_B/\mathcal{E}_p$ and $\alpha_E = U_E/\mathcal{E}_p$ are the respective magnetic and electric equipartition parameters. Here the energy in the magnetic fields  ($U_B = \int B^2 dV$) and electric fields ($U_E = \int E^2 dV$) is normalized to the initial total kinetic energy in the system $\mathcal{E}_p =\sum_\alpha \int n_{0\alpha} m_e v_{\rm{fl}}^2/2 dV$, summing over each species $\alpha$; in our case the number of species is $N_{SP} =2,(e^-,e^+)$. Here $m_e$, $n_{0\alpha}$, and $v_{\rm{fl}}$, are the respective mass, density, and velocity of the species, and $V$ is the total volume of the two slabs. Above quantities will be used demonstrating the slowdown process that occurs during the interaction of two plasma slabs. 

\section{Parametric estimate of bounds on Dark-EM}
Based on our simulation findings and astrophysical observations such as the
Bullet Cluster~\cite{DM25}, Abell 520~\cite{DM21}, and other galaxy cluster
collisions~\cite{Harvey2014}, we placed a very strong constraint on the parameter
space ($\alpha_D, m_D$) for DM particles. Here $\alpha_D \equiv e_D^2/\hbar c$
and $m_D$ are the respective dark electromagnetic coupling constant and mass,
and $e_D$ is the dark electromagnetic charge of the particle. 

Our numerical simulations reported in Ref~\cite{Shukla2022} showed that the
self-interaction of two cold $(v_{\rm{fl}} \geq v_{th})$ plasma clouds leads to
the generation of the Weibel and oblique instabilities, which deflect particle
trajectories such that the particles acquire transverse momentum while losing
forward momentum.  We showed that for a typical plasma slab length larger than $L
\approx 10 v_{\rm{fl}}/\Gamma_{\rm{W}}$, a significant slowdown occurs.  In the
case of a hot $(v_{\rm{fl}} \approx v_{\rm{th}})$ plasma cloud, we observed a slowdown caused solely by the Weibel instability.  

When it comes to a dark matter plasma, this result implies that for
sufficiently large $L\omega_{pD}/c$ the DM slab slows down due to the Weibel
and oblique instabilities, where $\omega_{pD} = \sqrt{4\pi\rho_D\alpha_D \hbar
c/m_D^2}$ is the dark plasma frequency.  For simplicity, we have considered
symmetric dark matter slabs, but in principle similar results should occur as
long as the less dense slab (defining $\omega_{pD}$) passes through a slab with
significant $L\omega_{pD}/c$.  In any case, the slowdown of the two
counterpropagating DM plasma slabs contradicts observations unless the
interaction is sufficiently weak.  Given some conservative estimates for $L
\approx 100\,\rm{kpc}$, $v_{\rm{fl}} \approx 0.1\,c$, $m_D \approx
1\,\rm{TeV}$, and the mass density $\rho_D \approx 0.01\,\rm{GeV/cm}^{3}$
\cite{DM25, DM8, DM10}, we can recast our limit of a maximum $L$ to avoid
slowdown to a limit on the coupling strength ($\alpha_D \ll 1)$.

%\emph{Limit based on the Weibel instability\,:\textemdash}
First, we estimate the coupling constant based on the Weibel growth. We found a significant slowdown occurs if the length $L > 10\,\Gamma_{\rm{W}}^{-1} v_{\rm{fl}}$, or a collision without significant slowdown requires 
\begin{align}
	\frac { L \Gamma_{\rm{W}} } {10\,v_{\rm{fl}}} = \frac { L } { 10\,v_{\rm{fl}}} \omega_{pD} \frac { v_{\rm{fl}}} { c } < 1
\end{align}
Recast in terms of $\alpha_{D}$ this is equivalent to the constraint 
\begin{align}
	\alpha_D < L^{-2} \rho_D^{-1} m_D^2 \left( \frac {100\,c } { 4\pi \hbar } \right) 
\end{align}

Using our estimates for the parameters, we find a quantitative limit on $\alpha_{ D }$ in engineering form
\begin{align}
\label{Weibelconstraint}
		\left.\begin{array} {c} { \alpha_D < 4.2355\times 10^{-25} \left( \frac {L} { 100\,\rm{kpc} } \right)^{-2} \left( \frac { \rho_D} { 0.01\,\rm{GeV/cm}^3} \right)^{-1} } \\ { \times \left( \frac {m_D} { 1\,\rm{TeV} } \right)^2 } \end{array} \right.
\end{align}

%\emph{Limit based on the oblique instability\,:\textemdash}
Similarly, we can estimate the coupling constant due to the oblique instability. The limit on $L$ for the oblique instability is
\begin{align}
			\frac { L \Gamma_{\rm{TS}} } { 10v_{\rm{fl}} } = \frac { L } { v_{\rm{fl}} } \omega_{pD} < 1,
\end{align}
which recast in terms of $\alpha_D$ is
\begin{align}
	\alpha _ { D } < L ^ { - 2} \frac{v_{\rm{fl}}^2}{c^2} \rho _ { D } ^ { - 1} m _ { D } ^ { 2} \left( \frac { 100 c} { 4\pi \hbar } \right)
\end{align}
and in engineering form is
\begin{align}
\label{obliqueconstraint}
	\left.\begin{array} { c } { \alpha _ { D } < 4.2355\times 10^ { - 27} \left( \frac { L } { 100\,\rm{kpc}} \right) ^ { - 2} \left( \frac { \rho_D } { 0.01\,\rm{GeV/cm}^3 } \right) ^ { - 1} } \\ { \times \left( \frac { m _ { D } } { 1\,\rm{TeV}} \right) ^ { 2} \left( \frac { v_{\rm{fl}} } { 0.1\,c } \right) ^ { 2} } \end{array} \right.
\end{align}
%--fig7
\begin{figure}[htp!]
	\begin{centering}	
		\includegraphics[width=0.4\textwidth]{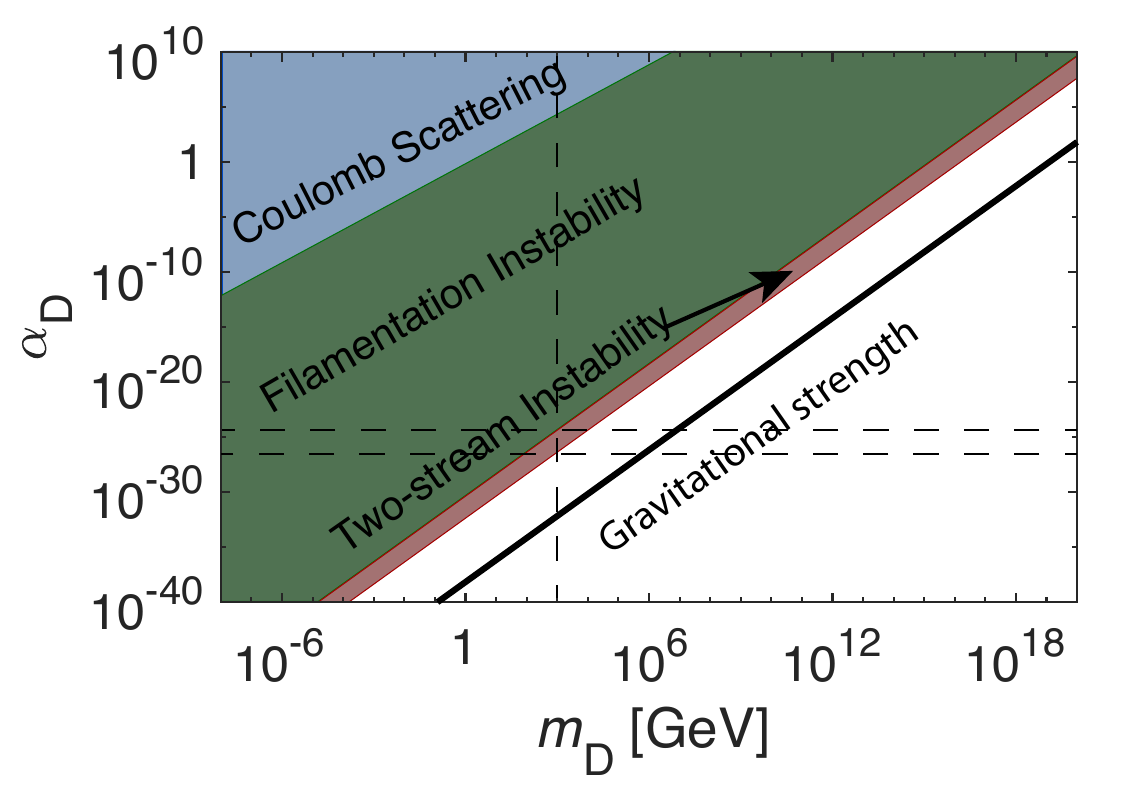}
	\end{centering}
		\caption{Constraints on the dark electromagnetic coupling constant $\alpha_{ D }$ as a function of $m_D$ based on the observations from the Bullet Cluster, assuming $\rho_D = 0.01\,\rm{GeV/cm}^3$,  $L = 100\,\rm{kpc}$,  $v_{\rm{fl}} = 0.1\,c$, and $\ln(\Lambda_C) = 35$. $\alpha_{ D }$ must lie below the region in blue due to the expected slowdown from Coulomb scattering (Eq.\,\ref{coulombconstraint}), below the region in green from the current filamentation instability (Eq.\,\ref{Weibelconstraint}), and below the region in red from the two-stream/oblique instability (Eq.\,\ref{obliqueconstraint}). For reference, the dark electromagnetic coupling forces will be equal to the gravitational forces along the bottom black line (Eq.\,\ref{gravitylimit}). The reverence value used in our equations, $m_D = 1\,\rm{TeV}$, is highlighted with a dashed line, along with the limits based on this value from Eqs.\,\ref{Weibelconstraint} and \ref{obliqueconstraint}}\label{DM1Fig9}
\end{figure}

 The previous bound on $\alpha_D$ that was established assuming a slowdown caused by Coulomb scattering \cite{DM21} is, in contrast, much weaker. For this limit the collision frequency
\begin{equation}
\nu = 
        \omega_{pD}^{4}
	\frac{m_D}{\rho_D}\frac{\ln(\Lambda_C)}{2 \pi v_{\rm{fl}}^3}
\end{equation}
where $\ln(\Lambda_C)$ is the Coulomb logarithm, is comparable to the crossing time; $\nu L/v_{\rm{fl}} = 1$, and in engineering form

\begin{eqnarray}
\label{coulombconstraint}
	\alpha_D &<& 1.9454 \times 10^4 \left(\frac{L}{100\,\rm{kpc}}\right)^{-1/2}
	\left(\frac{\rho_D}{0.01\,\rm{GeV/cm}^3}\right)^{-1/2} \nonumber\\
	&\times&\left(\frac{m_D}{1\,\rm{TeV}}\right)^{3/2}
	\left(\frac{v_{\rm{fl}}}{0.1 c}\right)^2
\left(\frac{\ln(\Lambda_C)}{35}\right)^{-1/2}.
\end{eqnarray}

Fig.\,\ref{DM1Fig9} shows the two limits from Eqs.\,(\ref{Weibelconstraint},\ref{obliqueconstraint}) in contrast to Eq.\,\ref{coulombconstraint}.
In the blue region, a slowdown due to Coulomb
scattering is expected.  In the green region, a slowdown due to the Weibel and oblique instabilities is expected.  In the red region, a moderate slowdown due to the oblique instability is expected if the plasma is cold ($v_{\rm{th}} \ll
v_{\rm{fl}}$).  Only the white region (and in some cases the red regime) is
consistent with observations, where no slowdown is expected. 

This establishes a strong upper bound on
the strength of a dark electromagnetic self-interaction.
To emphasize how strong the bound
on $\alpha_D$ this establishes, we also include a line in Fig.\,\ref{DM1Fig9} corresponding to when the strength of
the dark electromagnetic force between two dark matter particles is equal in magnitude to the gravitational force. To ignore the effects of gravity, the coupling constant needs to be sufficiently strong. The coupling constant that satisfies this constraint is
\begin{equation}
\label{gravitationalpha}
        \alpha_D > \frac{G}{\hbar c} m_D^2,
\end{equation}
where $G$ is the gravitational constant.
In engineering form this is 
\begin{equation}
\label{gravitylimit}
        \alpha_D > 6.7086 \times 10^{-33}
        \left(\frac{m_D}{1\,\rm{TeV}}\right)^2.
\end{equation}

We can show that the upper bound is statistically significant by
estimating an effective drag force (due to the dark electromagnetic fields) on the dark matter based on two values measured from the main simulation $R_1$ from Ref.~\cite{Shukla2022}.
We can express the estimated drag force in terms of the magnetic isotropization time $\Delta t_{\alpha_B} = 105.4\,\omega_{Dp}^{-1}$, and the slowdown of the flow during the collision $(v_{\rm{fl}}-v_{\rm{init}})/v_{\rm{fl}} = 0.8556$: 
\begin{equation}
	D_D = m_D \frac{v_{\rm{fl}}-v_{\rm{init}}}{\Delta t_{\alpha_B}}.
\end{equation}
Equating this with the drag force from Ref.~\cite{Harvey2014} $D_D = \rho_D m_D
v_{\rm fl}^2 \sigma/m$, we can obtain an effective dark matter cross section
$\sigma/m = 5.9~\text{cm}^2/\text{g}$. Here we have used our estimates of $\rho_D$ and $L$ and assumed
a regime with significant slowdown $\omega_{pD} = 10\,c/L$.
	As the typical measured value for $\sigma/m = 0 \pm 1~\text{cm}^2/\text{g}$~\cite{Harvey2014},
the slowdown in this region is statistically inconsistent with observation.

Larger
$\alpha_D$ particles could potentially be explained by more complex theories
which include an ionization fraction, where only a select few particles have
this larger interaction \cite{DM21}.

\section{Conclusion}
In conclusion, by comparing {\it ab initio} kinetic plasma simulations of the
interpenetration of two $e^-,e^+$ plasma slabs with astronomical observations,
we establish a strong upper bound on the strength of the dark electromagnetic
self-interaction $\alpha_{D}$  ($\alpha_{D} \approx 4 \times 10^{-25} \ll 1$, for $m_D = 1\,\rm{TeV}$).
This bound assumes the most basic unbroken U(1) gauge interaction of the dark
plasma.  As we have shown that the allowed parameter space of self-interacting DM particles, based on our analysis, exhibits significantly stronger bounds on
$\alpha_{D}$ than previously established, the electromagnetic interaction
between DM is considerably restricted.

It should be noted that the limit put on the self-interaction coupling strength is based on the assumptions that the plasma slabs are step functions in density and that there is no background dark magnetic field. 

A step function taking the average density and length gives a good estimate of the interaction between two dark matter halos with an arbitrary smooth distribution. 
However, one can make an even more conservative estimate on the coupling constant by only considering a small section of the interacting halos where the density gradient can be neglected. A recent complementary work just published~\cite{Derocco2024} concurrent with this publication uses particle-in-cell simulations to find a limit equivalent to $\alpha_D < 7.8\times 10^{-22} (m_D/1 \rm{TeV})^2$, which is comparable to Eq.\,~\ref{Weibelconstraint} if we assume $L$ is approximately $1\%$ smaller, so that the gradient can be neglected.
The spatial-temporal effects of the current filamentation instability present when the slabs are not initially superimposed causes a weaker growth before the purely temporal growth sets in \cite{Pathak2015}. Therefore, our constraint gives a more conservative estimate on the limit of the coupling.

Finally, we note that a uniform background dark magnetic field could lead to a suppression of the instability and slowdown. Although a magnetic field perpendicular to drift does not affect the growth rate \cite{Pokotelov2012},
a sufficiently strong magnetic field along the drift can suppress the instability \cite{Molvig1975,Li2005}. The criteria for suppression is equivalent to the regime where the system is stable to the firehose instability~\cite{Parker1958}, an instability derived in the magnetized regime also driven by the temperature anisotropy. We conjecture that a misalignment of the temperature and density gradients of the dark matter halos can lead to a large-scale dark magnetic field competing with the Weibel/current filamentation instability\cite{Schoeffler2014,Schoeffler2016}, with a plasma beta $\beta \sim L^2\omega_{pD}^2/c^2$. However, even in the hot regime $(v_{\rm{fl}} \approx v_{\rm{th}})$, this would be too weak to suppress the instability. Nevertheless, the limit on the dark matter coupling relies on the assumption that there are no dark currents, which can sustain a strong dark magnetic field parallel to the relative drift between the two DM halos, and thus no background field is considered.

Although we have had several fruitful discussions with experts in dark matter, the authors are primarily experts in the plasma physics aspects of this paper.
We hope that our results may be useful for the dark matter community and that the constraints found in this paper can be further verified statistically using observational measurements and uncertainties.
	%However, a slowdown is less efficient in this case. 

\begin{acknowledgments}
This study was completed as a part of N. Shukla's doctoral research~\citep{Shukla2019thesis,ShuklaAPS} under the supervision of K. Schoeffler. We thank Dr. J. Mardon and Dr. B. Feldstein who originally proposed the scientific problem, contributed to many fruitful discussions during the initial stage of this work, and helped give us a grounding as to the relevance of this work for the field of study on dark matter. We also thank Prof. M. Heikinheimo and Dr. P. Agrawal for useful discussions and feedback. N. Shukla and K. Schoeffler contributed equally to this work. This work was partially supported by the European Research Council (ERC-2015-InPairs-695088). Simulations were performed at the IST cluster (Lisbon, Portugal) and on the supercomputer Marconi (Cineca, Italy) in the framework of HPC-Europa3 Transnational Access programme.

\end{acknowledgments}

\appendix

\section{Particle-in-cell (PIC) method}
Particle-in-cell (PIC) codes are simulation tools that enable the modeling of complex kinetic and relativistic plasma phenomena on the smallest time scale i.e. electron time scales. The simulations were performed with the fully relativistic, massively parallel, PIC code OSIRIS \cite{R1, R2, R3}. In the PIC method,  the relativistic Lorentz force is used to calculate the motion of the plasma particles by interpolating the grid fields to the position of the particle. Current density is deposited onto the grid and used to advance the electric and magnetic fields through Maxwell's equations discretized using a finite-difference scheme. Each particle is pushed to a new position and momentum via self-consistently calculated fields. The method is based on a multi-step process and is second-order accurate in time.

%\bibliographystyle{plain}
\bibliography{DM_ref}

\end{document}